# Quantum-Annealing-Inspired Algorithms for Track Reconstruction at High-Energy Colliders

Hideki Okawa[1] · Qing-Guo Zeng[2,3,4,5] · Xian-Zhe Tao[2,3,4,5] · Man-Hong Yung[2,3,4,5]

**Abstract**
Charged particle reconstruction or track reconstruction is one of the most crucial components of pattern recognition in high-energy collider physics. It is known to entail enormous consumption of computing resources, especially when the particle multiplicity is high, which will be the conditions at future colliders, such as the High Luminosity Large Hadron Collider and Super Proton–Proton Collider. Track reconstruction can be formulated as a quadratic unconstrained binary optimization (QUBO) problem, for which various quantum algorithms have been investigated and evaluated with both a quantum simulator and hardware. Simulated bifurcation algorithms are a set of quantum-annealing-inspired algorithms, known to be serious competitors to other Ising machines. In this study, we show that simulated bifurcation algorithms can be employed to solve the particle tracking problem. The simulated bifurcation algorithms run on classical computers and are suitable for parallel processing and usage of graphical processing units, and they can handle significantly large amounts of data at high speed. These algorithms exhibit reconstruction efficiency and purity comparable to or sometimes improved over those of simulated annealing, but the running time can be reduced by as much as four orders of magnitude. These results suggest that QUBO models together with quantum-annealing-inspired algorithms are valuable for current and future particle tracking problems.

**Keyword** Quantum algorithm · Quantum annealing · Simulated bifurcation · Pattern recognition · HEP particle tracking · Combinatorial optimization

✉ Hideki Okawa
okawa@ihep.ac.cn

1 Institute of High Energy Physics, Chinese Academy of Sciences, 19B Yuquan Road, Shijingshan 100049, Beijing, China

2 Shenzhen Institute for Quantum Science and Engineering, Southern University of Science and Technology, 1088 Xueyuan Road, Shenzhen 518055, Guangdong, China

3 International Quantum Academy, 1088 Xueyuan Road, Shenzhen 518048, Guangdong, China

4 Guangdong Provincial Key Laboratory of Quantum Science and Engineering, Southern University of Science and Technology, 1088 Xueyuan Road, Shenzhen 518055, Guangdong, China

5 Shenzhen Key Laboratory of Quantum Science and Engineering, Southern University of Science and Technology, 1088 Xueyuan Road, Shenzhen 518055, Guangdong, China

## Introduction

High-energy colliders have played important roles in particle physics, including in the discovery of new particles and precise measurement of their interactions and properties. The Higgs boson, which was the last missing piece of the Standard Model, was discovered in 2012 at the Large Hadron Collider (LHC) at the European Organization for Nuclear Research (CERN) during the ATLAS and CMS experiments [1, 2]. The LHC has been successfully operating since 2009 and will be upgraded to the High Luminosity LHC (HL-LHC) [3] with increased luminosity compared to the current LHC Run 3 operation by several factors, which is scheduled to start after 2029. The event rate will also increase by a factor of 10 because of the trigger upgrade. Over 10 times the data recorded at the LHC will be collected at the HL-LHC, leading to approximately 180 million Higgs bosons being produced. This will allow us to measure the Higgs sector with unprecedented precision. With such an enormous amount of data collected and the increase in the data size by a factor of four to five due to the new detector, the LHC will shift from the current peta-byte data operation to the "exa-byte" era. The HL-LHC is expected to be followed by future colliders such as the Circular Electron Positron Collider (CEPC) [4–6] and Super Proton–Proton Collider (SppC) [7, 8] to be hosted in China as well as similar projects under consideration worldwide.

Charged particle reconstruction, track reconstruction or tracking in short is one of the most crucial components of reconstruction and is the reconstruction task with the highest CPU consumption at the LHC. It is a pattern recognition procedure used to recover charge particle trajectories from hits in the inner-most detector, which is fully composed of silicon layers at the HL-LHC, for example. As the trajectories are bent by the solenoid magnetic field, measuring their curvature provides the particle momenta. The number of reconstructed tracks per event at the LHC has been on the order of hundreds but will increase by an order of magnitude at the HL-LHC, reaching approximately ten thousand tracks at most. The required computing time exponentially increases with the luminosity [9–11], and various innovations are urgently needed to overcome this challenge.

The Kalman filter [12] has been used as a standard algorithm at the LHC and is also implemented in an open-source project called A Common Tracking Software (ACTS) [13]. The Kalman filter initiates seeding from the inner layers of the tracking detector, and the tracks are extrapolated to find the next hit in the outer layers. The track fit is iterated to find the best quality among the hit combinations. Owing to its highly efficient track reconstruction and very low misidentification rate, the Kalman filter remains as the key component for tracking. However, owing to its iterative process, the computing time exponentially increases with the track multiplicity.

To address this challenge, machine learning methods such as graph neural networks (GNNs) are actively being investigated at the LHC [14, 15] and via the Beijing Spectrometer (BES) III [16] at the tau-lepton and charm quark factory in China. The nodes of the graphs represent hits in the silicon detector, and the edges represent segments obtained by connecting these silicon hits. The GNN-based approaches provide track reconstruction performance comparable to that of the Kalman filter, but the computing time scales approximately linearly.

The application of quantum computing and algorithms is yet another trend of innovative approaches being investigated to address track reconstruction in dense conditions. The first such studies [17, 18] utilized a quantum annealing computer, considering track reconstruction as a quadratic unconstrained binary optimization (QUBO) problem. Modeling track reconstruction as a QUBO problem and solving it via simulated annealing actually go back to the time of the ALEPH experiment [19]. However, the experimental conditions at that time of low track multiplicity and a nondemanding computing load significantly differ from those of the current LHC and future HL-LHC; thus, several modifications were necessary, such as those implemented in Refs. [17, 18]. In recent years, several groups have investigated the possibility of using quantum gate simulators and computers instead [20–25]. With quantum gates, a QUBO can be mapped to an Ising Hamiltonian and be solved via the Variational Quantum Eigensolver (VQE), the Quantum Approximate Optimization Algorithm (QAOA), or similar variants. Recently, introducing GNN tracking into quantum gate computers [26] or quantum annealing machines [27] has also been proposed. There is also a conceptual study categorizing track reconstruction into four routines and demonstrating how quantum advantages can be achieved [28].

## Methodology

Track reconstruction can be regarded as an Ising problem. It is a nondeterministic polynomial time (NP) complete problem, in which the solution candidates exponentially diverge with the problem size. Machines dedicated to solving this class of problems are called Ising machines. The quantum annealer by D-Wave [29] based on the quantum annealing concept [30] and the coherent Ising machine (CIM) [31] are such examples.

In 2016, a new quantum computer modifying CIM, the quantum bifurcation machine (QbM) [32, 33], was proposed. The classical version of QbM, the classical bifurcation machine (CbM), was also introduced in parallel [32, 33]. This classical algorithm no longer has a foundation in the quantum adiabatic theorem but can still find quasi-optimal solutions. The original equation of motion of CbM was then modified to speed up the numerical simulation, and this variant is called simulated bifurcation (SB) [34]. This quantum-annealing-inspired algorithm (QAIA) is suitable for parallel computing and expected to outperform CIM. The details of SB, the simulated annealing (SA) [35] approach considered the benchmark in this study, and how track reconstruction is formulated as an Ising or QUBO problem are provided below. The simulated CIM [36] showed visibly degraded performance for track reconstruction in terms of both the predicted minimum Ising energy and the speed to obtain an answer; thus, it is not presented in this work.

### Simulated Bifurcation

The first SB proposed, adiabatic SB (aSB), is formulated as:

$$\dot{x}_i = \frac{\partial H_{\text{aSB}}}{\partial y_i} = a_0 y_i, \tag{1}$$

$$\dot{y}_i = -\frac{\partial H_{\text{aSB}}}{\partial x_i}$$
$$= -\left[x_i^2 + a_0 - a(t)\right]x_i + c_0 \sum_{j=1}^{N} J_{i,j} x_j, \tag{2}$$

$$H_{\text{aSB}} = \frac{a_0}{2} \sum_{i=1}^{N} y_i^2 + V_{\text{aSB}}, \tag{3}$$

$$V_{\text{aSB}} = \sum_{i=1}^{N} \left(\frac{x_i^4}{4} + \frac{a_0 - a(t)}{2} x_i^2\right) - \frac{c_0}{2} \sum_{i=1}^{N} \sum_{j=1}^{N} J_{i,j} x_i x_j, \tag{4}$$

where $x_i$ and $y_i$ are, respectively, the position and momentum of a particle corresponding to the $i$th spin, $a(t)$ is a time-dependent control parameter that monotonically increases from zero, $a_0$ and $c_0$ are positive constants, $H_{\text{aSB}}$ is the Hamiltonian, and $V_{\text{aSB}}$ is the potential energy in aSB. The symplectic Euler method is adopted for the numerical computation [37]. All the positions and momenta can be instantaneously updated; thus, this method is suitable for parallel computation, which is a notable difference from simulated annealing.

Despite the abovementioned appealing characteristics, the precision of aSB was found to be problematic, so a few variants of the algorithm were proposed [38]. The degradation of the aSB performance mostly originates from the fact that it treats the discrete values of the spins as the continuous variables $x_i$. The ballistic SB (bSB) algorithm introduces complete inelastic scattering walls at $|x_i| > 1$, meaning that $x_i$ is forced to be $x_i = \pm 1$ and $y_i = 0$ when the particle passes the threshold. The equation of motion is formulated as follows:

$$\dot{x}_i = \frac{\partial H_{\text{bSB}}}{\partial y_i} = a_0 y_i, \tag{5}$$

$$\dot{y}_i = -\frac{\partial H_{\text{bSB}}}{\partial x_i}$$
$$= -\left[a_0 - a(t)\right]x_i + c_0 \sum_{j=1}^{N} J_{i,j} x_j, \tag{6}$$

$$H_{\text{bSB}} = \frac{a_0}{2} \sum_{i=1}^{N} y_i^2 + V_{\text{aSB}}, \tag{7}$$

$$V_{\text{bSB}} = \frac{a_0 - a(t)}{2} \sum_{i=1}^{N} x_i^2 - \frac{c_0}{2} \sum_{i=1}^{N} \sum_{j=1}^{N} J_{i,j} x_i x_j \quad \text{(when } |x_i| \le 1 \text{ for all } x_i\text{)}$$
$$= \infty \quad \text{(otherwise),} \tag{8}$$

As a potential wall is introduced at $x_i = 1$, the quartic term is removed to avoid divergence. In addition to showing better precision, bSB is approximately 12 times faster than aSB when implemented in a field programmable gate array (FPGA) [38].

The discrete SB (dSB) algorithm partially introduces discretization of the position into the equation of motion, replacing $x_i$ with $\text{sgn}(x_i)$. This approach is expected to reduce the analog error. More interestingly, due to the violation of energy conservation arising from the discretization, it shows a pseudo-quantum-tunneling behavior. Its equation of motion is written as follows:

$$\dot{x}_i = \frac{\partial H_{\text{dSB}}}{\partial y_i} = a_0 y_i, \tag{9}$$

$$\dot{y}_i = -\frac{\partial H_{\text{dSB}}}{\partial x_i}$$
$$= -\left[a_0 - a(t)\right]x_i + c_0 \sum_{j=1}^{N} J_{i,j} \text{sgn}(x_j), \tag{10}$$

$$H_{\text{dSB}} = \frac{a_0}{2} \sum_{i=1}^{N} y_i^2 + V_{\text{aSB}}, \tag{11}$$

$$V_{\text{dSB}} = \frac{a_0 - a(t)}{2} \sum_{i=1}^{N} x_i^2 - \frac{c_0}{2} \sum_{i=1}^{N} \sum_{j=1}^{N} J_{i,j} x_i \, \text{sgn}(x_j)$$
$$\text{(when } |x_i| \le 1 \text{ for all } x_i\text{)}$$
$$= \infty \quad \text{(otherwise),} \tag{12}$$

where the abovementioned partial discretization is introduced by replacing some $x_i$ with $\text{sgn}(x_i)$.

Another set of variants introducing heat fluctuations [39] is not considered in this study, as these variants do not significantly improve the results. The SB algorithms are implemented in Huawei MindSpore Quantum [40] by one of the authors and are adopted in this study. aSB is already significantly outperformed by the other variants and is not considered. A recent application of SB algorithms to drug design is documented in Ref. [41].

## Simulated Annealing

In this work, the simulated annealing algorithm implemented in D-Wave Neal [42] is considered the baseline for comparison, as its performance in tracking is already known and has been presented in a previous study [17, 43]. D-Wave Neal is based on the method proposed in Ref. [35].

It uses random moves in the solution space to search for the global minimum. Analogous to finding ground states of matter, a "temperature" is introduced to first "melt" a substance at a high temperature, and then, the temperature is slowly lowered until the system "freezes", similar to growing a crystal from a melt. A simple algorithm proposed in Ref. [44] is employed for the iteration, in which a small random displacement is applied to the original spin configuration, and the energy difference $\Delta E$ is computed. If $\Delta E < 0$, then the new configuration is accepted as the new temporary prediction, and the algorithm moves on to the next iteration. If $\Delta E > 0$, then the new configuration is accepted only probabilistically according to the Boltzmann factor: $P(\Delta E) = \exp(-\Delta E/k_B T)$.

Another QUBO solver, qbsolv [45], provided by D-Wave, uses a sub-QUBO method, in which the original QUBO matrix is partitioned into pieces. These smaller subproblems are then solved using a classical solver running the tabu algorithm [46]. Sub-QUBO methods are mainly motivated by implementation in actual quantum annealing hardware, which has limitations in terms of the problem size due to the available number of qubits. The speed of qbsolv is more than two orders of magnitude slower than that of Neal, with no gain in performance in track reconstruction [17, 43]. Thus, it is not considered in our study.

## QUBO Formulation

To reconstruct the tracks, hits in the silicon detectors are connected, starting from doublets, i.e., segments of two silicon hits. Then, triplets, i.e., segments of three silicon hits, are created by connecting the doublets. Triplets are connected to reconstruct the final tracks, by evaluating the consistency of the triplet momenta. The following formulation of a QUBO considered in Refs. [17, 23, 24, 43, 47] is adopted:

$$O(a, b, T) = \sum_{i=1}^{N} a_i T_i + \sum_{i=1}^{N} \sum_{j<i}^{N} b_{ij} T_i T_j, \quad (13)$$

where $N$ is the number of triplets, $T_i$ and $T_j$ correspond to the triplets and take the value of either zero or one, $a_i$ are the bias weights used to evaluate the quality of the triplets, and $b_{ij}$ are the coefficients that quantify the compatibility of two triplets ($b_{ij} = 0$ if there is no shared hit, $= 1$ if there is any conflict, and $= -S_{ij}$ if two hits are shared between the triplets). The coefficients $a_i$ and $b_{ij}$ are unitless; thus, the objective QUBO function $O(a, b, T)$ gives unitless energy. The coefficients $-S_{ij}$ quantify the consistency of the two triplet momenta via [17]:

$$S_{ij} = \frac{1 - \frac{1}{2}(|\delta(q/p_{Ti}, q/p_{Tj})| + max(\delta\theta_i, \delta\theta_j))}{(1 + H_i + H_j)^2}, \quad (14)$$

where $\delta$ is the difference between the curvature $q/p_T$ or angle $\theta$ values of the two triplets and $H_i$ is the number of holes in the triplet. The bias weights $a_i$ have a significant effect on the Hamiltonian energy landscape and thus on track reconstruction, particularly its purity (defined in Sec. 5) and computation speed. They can be parameterized as:

$$a_i = \alpha \left(1 - e^{\frac{|d_0|}{\gamma}}\right) + \beta \left(1 - e^{\frac{|z_0|}{\lambda}}\right), \quad (15)$$

where $d_0$ and $z_0$ are the transverse and longitudinal displacements of the triplets from the primary vertex (the most significant proton–proton collision point in an event) and $\alpha, \beta, \gamma$ and $\lambda$ are tunable parameters, which are taken to be 0.5, 0.2, 1.0 and 0.5 based on an optimization in a previous study [43] for the same dataset, which is described in the next section.

## Dataset

An open-source dataset from the TrackML Challenge is used [49, 50]. The events contain a top quark pair generated with the Pythia8 [51] event generator overlaid with additional 200 proton–proton interactions compatible with the HL-LHC conditions. A generalized full silicon tracker called the TML detector, which mimics the inner detector component of the ATLAS experiment for the HL-LHC conditions, is considered. The detector consists of three detector volumes, which are cylindrical in the barrel (central) region with multiple disks in the end-cap (side) region. Each detector volume has several silicon layers, leading to 10 layers in total for the entire TML detector. However, the dataset is simplified by focusing on the barrel region of the detector, namely, removing the hits in the end-cap region. This approach is often adopted for tracking studies for the HL-LHC.

A fast detector simulation implemented in the ACTS software is applied to the charged particles traversing the silicon layers. A solenoid magnetic field of 2T with realistic inhomogeneous distortion of the field strength is considered. Material interactions such as multiple scattering, energy loss and hadronic interactions are parameterized in the simulation. Inefficiency in silicon sensors, false silicon hits and the production of secondary particles from the detector material are also considered. Figure 1 shows an event display from the highest particle multiplicity event in the dataset. This figure clearly demonstrates how dense the reconstructed tracks are under the HL-LHC conditions.

**Fig. 1** Display of the highest particle multiplicity event in the dataset. The green (red) lines indicate correctly (incorrectly) reconstructed tracks, whereas the blue lines are those not reconstructed. This display is generated with the hepqpr-qallse framework [48]

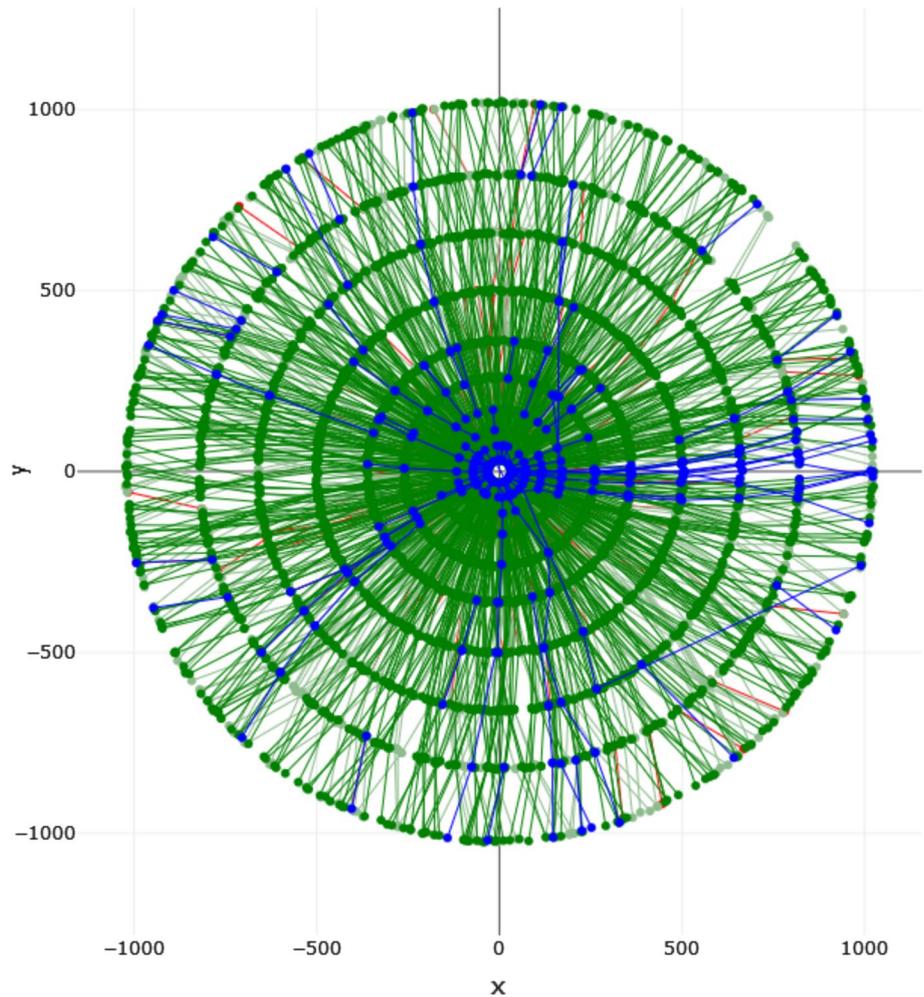

## QUBO Generation and Track Finding

Each event is converted to the QUBO format with the hepqpr-qallse framework [48]. Generating all potential doublets from the silicon hits results in a quadratic size and is not efficient. Thus, the initial doublets are generated with a Python library adapted from the online seeding code used in the ATLAS Trigger system [52]. The seeding code examines the length of the doublets and their angles to ensure that they are not significantly nonvertical. Triplets $T_i$ are formed by connecting the generated doublets but are discarded unless they satisfy the following requirements:

$$\begin{aligned} H_i &\leq 1, \\ |(q/p_\mathrm{T})_i| &\leq 8 \times 10^{-4}\ \mathrm{GeV}^{-1}, \\ \delta\theta_i &\leq 0.1\ \mathrm{rad}. \end{aligned} \quad (16)$$

Similarly, quadruplets $(T_i, T_j)$, i.e., segments of four silicon hits created from the triplets, are kept only if:

$$\begin{aligned} |\delta((q/p_\mathrm{T})_i, (q/p_\mathrm{T})_j)| &\leq 1 \times 10^{-4}\ \mathrm{GeV}^{-1}, \\ S_{i,j} &> 0.2. \end{aligned} \quad (17)$$

Furthermore, quadruplets must be able to form at least one track candidate with five hits or more. Finally, doublets and triplets that do not belong to any quadruplet are discarded.

By solving the QUBO, the binary variables $T_i$ can be used to determine which triplets should be kept. The selected triplets are converted back to doublets, and duplicates or unresolved conflicts are removed. Track candidates are formed from these final doublets, and those with fewer than four silicon hits are discarded (an example of such final track candidates is presented in Fig. 1).

## Results

First, the performance of the three QAIAs, i.e., bSB, dSB and D-Wave Neal, is evaluated based on the predicted minimum Ising energy and its stability over 50 shots. For bSB and dSB, the parameter $a_0$ in Eqs. (6) and (10) and the time step $\Delta t$ are set to 1, similar to the proposed values in Ref. [38]. Two approaches are tested for the parameter $c_0$: the former involves keeping it fixed, whereas the latter involves pursuing an automatic optimization scan. There is no significant difference in performance for the two cases; thus, the results for the fixed $c_0$ parameter case are presented from here on. As shown in Fig. 2, bSB provides the lowest predicted minimum energy with the smallest fluctuation regardless of the QUBO size, followed by D-Wave Neal. dSB tends to have slightly degraded energy prediction with larger fluctuations, although its impact on the final track reconstruction performance is not critical, as will be presented below.

The actual track reconstruction performance is evaluated in terms of the efficiency and purity, or the recall and precision in data science terminology. They are defined as follows:

$$\text{Efficiency (Recall)} = \frac{TP}{TP + FN}, \tag{18}$$

$$\text{Purity (Precision)} = \frac{TP}{TP + FP}, \tag{19}$$

where $TP$ represents the true positives, $FN$ represents the false negatives, and $FP$ represents the false positives. $TP$ corresponds to the number of reconstructed doublets matching the correct (true) doublets. $FN$ is the number of true doublets that are not reconstructed; thus, $TP + FN$ is simply the number of true doublets. *FP* is the number of reconstructed doublets that do not match the true doublets, namely, the "fake" doublets in high-energy physics terminology. Figure 3 shows the efficiency and purity for the three QAIAs. The statistical uncertainty and the root-mean-square of the fluctuations from 50 shots are added in quadrature and presented in the figure. The statistical uncertainty is a few times greater than the fluctuations from 50 shots; thus, the latter effect is not significant for all three algorithms. An excellent efficiency of above 95% is obtained for all the algorithms throughout the whole dataset up to the highest particle multiplicity of approximately 10000. The purity decreases with increasing particle multiplicity but remains above 84% for all events and above 95% for events with a particle multiplicity of less than 6000. dSB tends to have slightly lower performance, although mostly within the statistical uncertainty. An event display from the highest multiplicity event reconstructed with bSB is presented in Fig. 1.

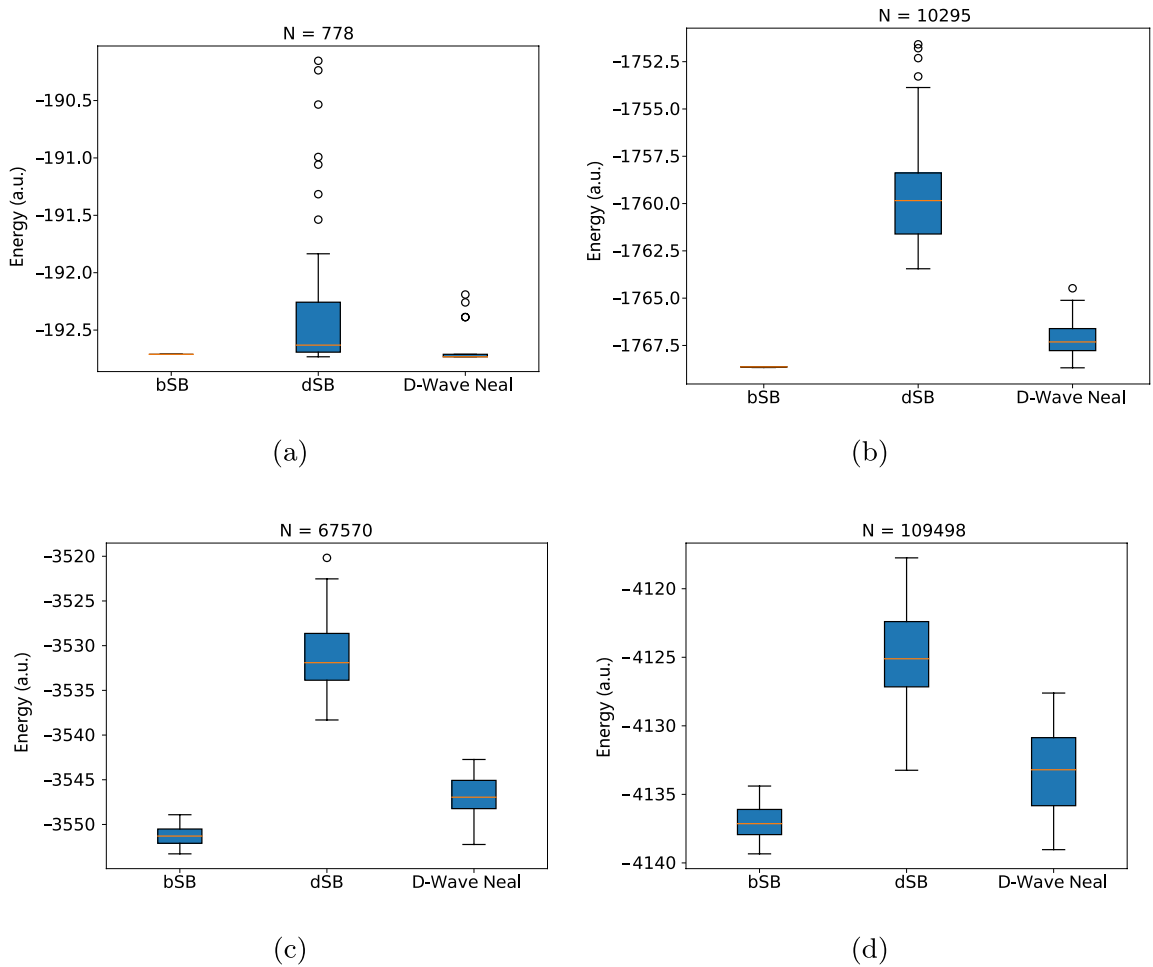

**Fig. 2** Minimum Ising energy estimated by various QAIAs. Four examples with a full QUBO size of 778×778 [409 particles] (**a**), 10,295×10,295 [4092 particles] (**b**), 67,570×67,570 [8187 particles] (**c**), and 109,498×109,498 [9435 particles] (**d**) are presented

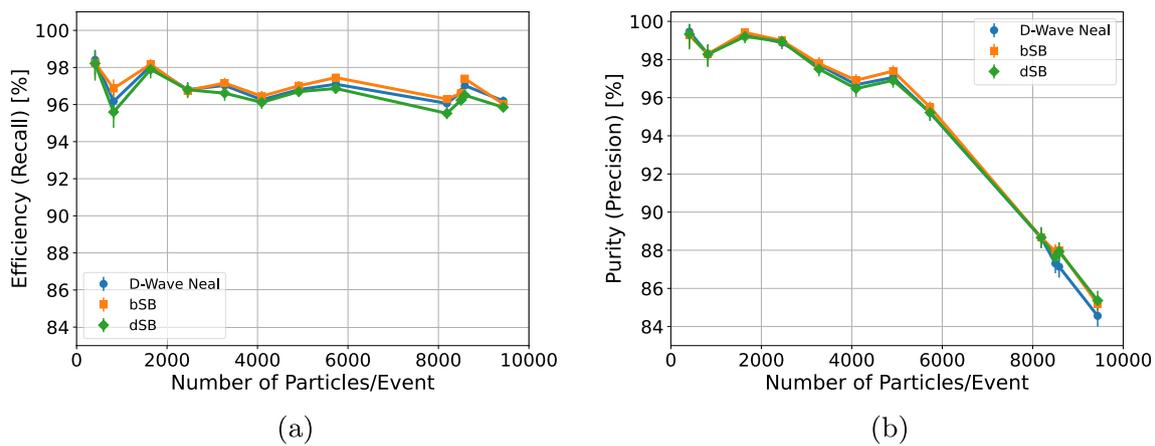

**Fig. 3** Efficiency (recall) (**a**) and purity (precision) (**b**) as a function of particle multiplicity evaluated for the three QAIAs

Finally, the execution time for each algorithm on an AMD EPYC 7642 CPU and an NVIDIA A100 GPU is evaluated. Figure 4 presents the evolution of Ising energies evaluated for the three QAIAs. Only one CPU or GPU is used for fair comparison with D-Wave Neal. The average of the 50 shots as well as the envelopes from the best and worst cases are shown in the figure. Usage of a GPU is generally not effective for simulated annealing algorithms because of the lack of parallelizability of the spin update; thus, the GPU option is not presented for D-Wave Neal.

More quantitative computation speeds are summarized in Table 1 using the time-to-target (TTT) metric [53]. The time-to-solution (TTS) metric [54, 55] is more standard when evaluating the Ising machine speed, but it requires the true ground state to be known. As the true ground state in our dataset is unknown, the TTT is adopted, where the target value is set to 99.9% of the lowest energy value obtained from our study. We adopt the time required to find the target value with 99% probability as the TTT. For small-sized QUBOs, bSB is one order of magnitude faster than D-Wave Neal. The impact of the GPU usage becomes significant for large-sized QUBOs, leading to a four-order-of-magnitude speed-up for the highest multiplicity event compared with D-Wave Neal. dSB cannot reach the target value with 99.9% probability for larger datasets. If the target requirement is loosened, then the speed of dSB is generally faster by a few factors with the CPU and faster by two to three orders of magnitude with the GPU than D-Wave Neal. The event with 8583 particles shows a TTT that does not follow the general trend with respect to the dataset size. This is because this event has a lower level of difficulty in being solved than the other events. The difficulty in solving a problem largely depends on the graph structure corresponding to the QUBO and its weight distribution.

## Conclusion and Outlook

Three sets of QAIAs are evaluated for track reconstruction formulated as a QUBO problem. bSB provides improved performance over the D-Wave Neal simulated annealers in all aspects: comparable or slightly higher performance in terms of the track reconstruction efficiency and purity and significant speed-up by four orders of magnitude at most are obtained when finding the minimum Ising energy with higher stability. This trend of improvement becomes more apparent as the dataset size increases, which is promising for future colliders, such as the HL-LHC and SppC. dSB shows a slightly degraded prediction of the minimum energy, but its impact on the efficiency and purity of the track reconstruction is not significant. dSB is generally two to three orders of magnitude faster than D-Wave Neal. Owing to their suitability for parallel processing and usage of cutting-edge computing resources, the SB algorithms are promising options for adoption in high-energy collider experiments. Notably, these are "quantum-inspired" algorithms running on classical computers, yet they are known to outperform quantum annealing for some existing problems [34]. Thus, they are not technologies for the future but are a serious option to be adopted for any ongoing collider experiments.

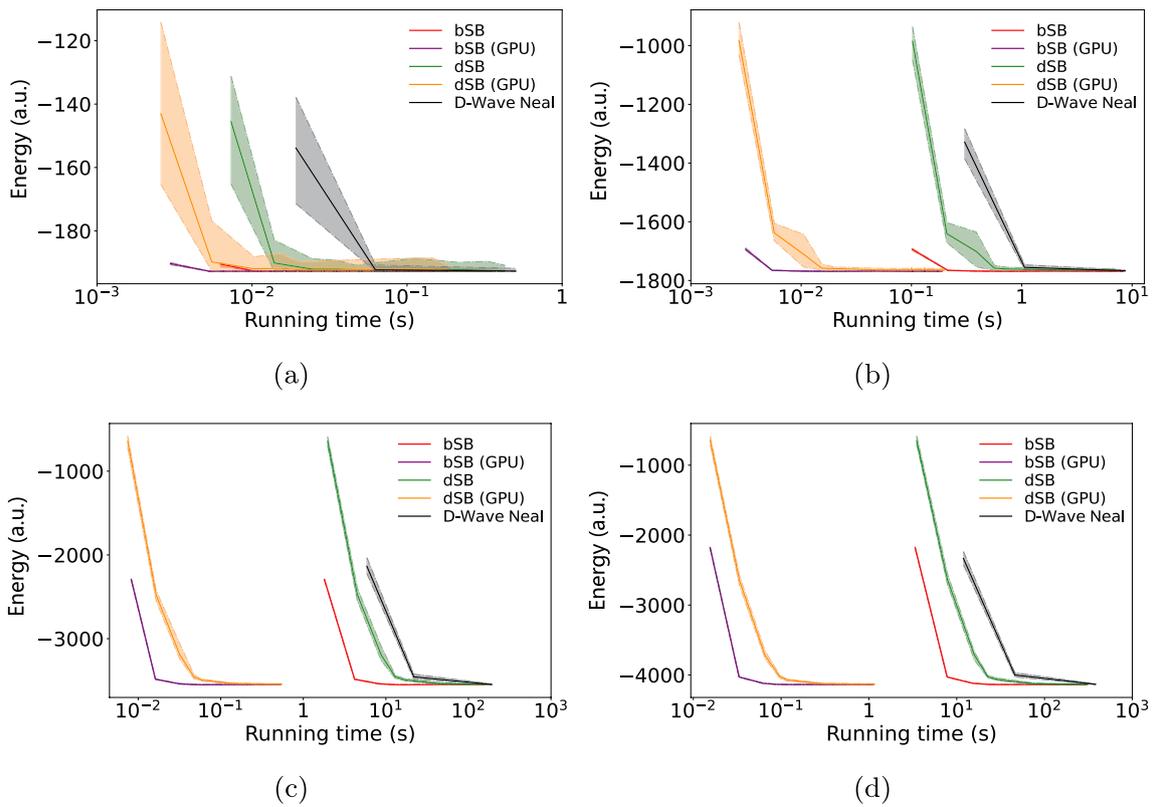

**Fig. 4** Evolution of Ising energies evaluated for the three QAIAs. Four examples with a full QUBO size of 778×778 [409 particles] (**a**), 10,295×10,295 [4092 particles] (**b**), 67,570×67,570 [8187 particles] (**c**), and 109,498×109,498 [9435 particles] (**d**) are presented. The solid lines indicate the average from the 50 shots, whereas the envelopes represent the best and worst cases among these shots

**Table 1** Time to target (TTT) for various particle multiplicities and QUBO sizes using bSB, dSB and D-Wave Neal. For bSB and dSB, both the CPU and GPU times are measured. The time is not shown if the prediction did not reach the target value

| Data information | | Time to target [s] | | | | |
|---|---|---|---|---|---|---|
| # of part. | QUBO | bSB | bSB (GPU) | dSB | dSB (GPU) | Neal |
| 409 | 778 | 0.007 | 0.021 | 0.032 | 0.092 | 0.060 |
| 818 | 1431 | 0.012 | 0.019 | 0.293 | 0.478 | 0.169 |
| 1637 | 2904 | 0.012 | 0.019 | 0.293 | 0.478 | 0.169 |
| 2456 | 4675 | 0.014 | 0.017 | – | – | 0.479 |
| 3274 | 6945 | 0.032 | 0.022 | – | – | 1.229 |
| 4092 | 10,295 | 0.005 | 0.022 | 0.015 | 0.065 | 0.030 |
| 4912 | 14,855 | 0.027 | 0.016 | – | – | 2.165 |
| 5730 | 22,022 | 0.109 | 0.042 | – | – | 3.853 |
| 8187 | 67,570 | 0.488 | 0.028 | – | – | 404.297 |
| 8500 | 78,812 | 1.899 | 0.108 | – | – | 785.732 |
| 8583 | 80,113 | 1.321 | 0.067 | – | – | 93.782 |
| 9435 | 109498 | 3.884 | 0.140 | – | – | 1366.808 |

Nevertheless, one thing to note is that the execution time for the QUBO model, namely, the doublet and triplet formation and preparation of a QUBO matrix from each given dataset, is not included in the results. This part of the procedure also takes a significant amount of time, but the execution is currently performed on a single CPU [48] and is definitely far from being optimized in terms of speed performance. It could be significantly improved with parallel processing and careful optimization. Its speed-up should not be a major bottleneck but is left for future studies.

**Acknowledgements** HO would like to thank Andreas Salzburger for his suggestion regarding the TrackML dataset and discussions about track reconstruction in general. HO is supported by the National Natural Science Foundation of China (NSFC) General Project (Grant No. 12075060) and the NSFC Basic Science Center Program for "Joint Research on High Energy Frontier Particle Physics" (Grant No. 12188102).

**Author contributions** H.O.: Concept, direction, conversion of the dataset into the quadratic unconstrained binary optimization (QUBO) format, performance evaluation of track reconstruction, prepared figures 1 and 3, wrote the main manuscript text, - Q.-G.Z.: Development and implementation of the quantum-annealing-inspired algorithms (QAIAs), Solving the QUBO problems with the QAIAs, evaluation of computing time, prepared figures 2, 4 and table 1, contributions to manuscript text, - X.-Z.T.: Solving the QUBO problems with the QAIAs, complementary studies to Q.-G.Z., - M.-H.Y.: supervision of Q.-G.Z. and X.-Z.T., contributions to manuscript text All authors reviewed the manuscript.

**Data availability** The source of data is provided within the manuscript by citing the relevant papers. The dataset URL is at https://www.kaggle.com/c/trackml-particle-identification/data

## Declarations

**Competing interests** The authors declare no competing interests.